\documentclass[fp,twocolumn]{jpsj3}
\usepackage{txfonts}
\usepackage{graphicx}

\title{Ferromagnetic Ising model on the hierarchical pentagon lattice}

\author{Takumi Oshima$^{1}$, and Tomotoshi Nishino$^2$\thanks{nishino@kobe-u.ac.jp}}
\inst{$^1$Department of Physics, Faculty of Science, 
Kobe University, Kobe 657-8501, Japan\\
$^2$Department of Physics, Graduate School of Science, 
Kobe University, Kobe 657-8501, Japan}

\abst{
Thermodynamic properties of the ferromagnetic Ising model on the hierarchical pentagon 
lattice is studied by means of the tensor network methods. The lattice consists of pentagons, 
where 3 or 4 of them meet at each vertex. Correlation functions on 
the surface of the system up to $n = 10$ layers are evaluated by means of the time evolving 
block decimation (TEBD) method, and the power low decay is observed in the high temperature 
region. The recursive structure of the lattice enables complemental numerical study for larger 
systems, by means of a variant of the corner transfer matrix renormalization group (CTMRG) 
method. Calculated spin expectation value shows that there is a mean-field type 
order-disorder transition at $T_1^{~} = 1.58$ on the surface of the system. On the other hand, 
the bulk part exhibits the transition at $T_2^{~} = 2.269$. Consistency of these calculated 
results is examined.
}

\begin{document}
\maketitle

\section{Introduction}

The order-disorder phase transition has been one of the central concern in modern 
statistical physics~\cite{PT}. The Ising model~\cite{Ising} has been extensively studied 
as a theoretical model of magnetic materials that consists of locally interacting molecular 
magnetic moments~\cite{Ising_review}. On the square lattice, presence of the phase 
transition was proven by Peierls~\cite{Peierls}, and the exact formula for the free energy 
in the thermodynamic limit was later obtained by Onsager~\cite{Onsager}. The concept 
of the renormalization group (RG) provides the unified picture on the singular behavior of 
thermodynamic functions around the phase transition point~\cite{Kadanoff1,Kadanoff2,Kogut}.
The nature of the second-order phase transition on the regular lattice that can be 
uniformly drawn on {\it the flat plane} is well understood from the view point of
the conformal field theory~\cite{Conformal}.

The Ising model on the Cayley tree lattices has been known as a reference model, 
where the partition function of the whole system can be easily obtained by taking spin 
configuration sum from the boundary sites~\cite{Baxter}. Although the corresponding free 
energy is an analytic function of the temperature $T$, those bulk spins deep inside the 
system, which are around the root of the tree, can posses finite spontaneous magnetic moment 
below the transition temperature, under the presence of infinitesimally weak external 
field~\cite{Runnels,Eggarter,Zittartz}. The transition is mean-field like, as it is explained 
from the self-consistent study on the Bethe lattice~\cite{Bethe,Baxter}. 

Similarly, on the hyperbolic $( 5, 4 )$ lattice, where four pentagons meet at each vertex, 
presence of the mean-field like phase transition in the bulk part 
of the system was confirmed numerically for the ferromagnetic Ising model by means of 
the corner transfer matrix renormalization group (CTMRG) 
method~\cite{1968,1978,CTMRG1,CTMRG2,Orus} adapted to the hyperbolic lattice 
structure~\cite{Ueda,Krcmar}. Since the $( 5, 4 )$ lattice is a regular lattice on {\it the 
negatively curved surface}, which has a finite curvature radius $R$ as the typical length 
scale, the bulk part of the system cannot be critical, where there is scale invariance~\cite{weak}. 
Thus the correlation length of the model (along the geodesics) is always finite, even at 
the bulk transition temperature~\cite{Iharagi}. It is naturally expected that ferromagnetic 
Ising models on the hyperbolic $( p, q )$ lattices, where $q$ numbers of $p$-gons meet 
at the lattice point, share the mean-field nature~\cite{Andrej_tri}.

Recently, Asaduzzaman et al performed the Monte Carlo simulation for the Ising model 
on the hyperbolic $( 3, 7 )$ lattice~\cite{Asaduzzaman}. From the numerical study 
on finite size systems, they confirmed presence of the power-law decay of the correlation 
function on the boundary of the system at any temperature. Okunishi and Takayanagi have 
rigorously shown the power-law decay along the boundary of the trivalent Cayley 
tree lattice~\cite{Okunishi}, which is the hyperbolic $( \infty, 3 )$ lattice, and reinterpreted the 
system from the view point of the Ads/CFT correspondence~\cite{Ads1,Ads2,Ads3,Ads4}.
One of the theoretical interest on the hyperbolic $( p, q )$ lattice is to confirm the presence, 
or absence, of the order-disorder transition at the system boundary.

\begin{figure}
\begin{center}
\includegraphics[width = 5.7 cm]{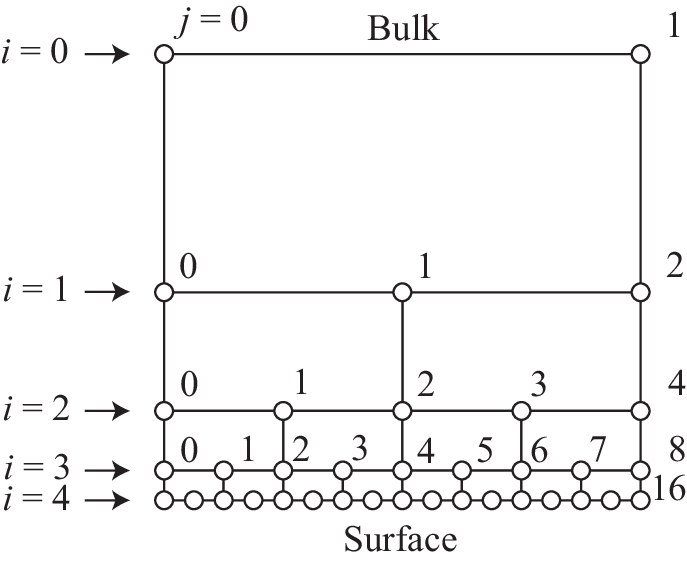}
\end{center}
\caption{Structure of the hierarchical pentagon lattice in the case $n = 4$. We regard
$\sigma_0^0$ and $\sigma_1^0$ at the top as the {\it bulk} spins, and those
$\sigma_j^4$ from $j = 0$ to $j = m(4) = 16$ at the bottom as the {\it surface} spins.}
\end{figure}

Motivated from these recent studies focused on the hyperbolic lattices, in this article 
we analyze the thermodynamic properties of the ferromagnetic Ising model on the 
hierarchical {\it pentagon} lattice shown in Fig.~1. Typically the case when there are 
$n = 4$ layers of horizontally aligned pentagons is drawn. It should be noted that 
all the pentagons are represented by the rectangular shape, so that the hierarchical 
lattice structure can be captured systematically. There are $2^n_{~} - 1$ pentagons 
in total. Three or four pentagons meet on each vertex, and the number is exceptionally 2 
at the system boundary. There is an Ising spin $\sigma_j^i = \pm 1$ shown by the open 
circle on each vertex, where the index $i$ specifies the row from the top 
$i = 0$ to the bottom $i = n$, and where $j$ specifies the horizontal location from the left 
$j = 0$ to the right $j = 2^i_{~}$. In order not to use nested index in the following equations, 
we introduce the notation $m( i ) = 2^i_{~}$, where $m( i ) + 1$  is the number of sites on 
the $i$-th row. 
The lattice has a geometrical analogy with the Cayley tree, in the sense that we obtain 
the binary tree by connecting the centers of vertically touching pentagons. 
Thus the upper boundary of the lattice with $i = 0$ corresponds to the root of the tree, 
and the lower boundary with $i = n$ corresponds to the leaves.
Considering the analogy, we regard 
$\sigma_0^0$ and $\sigma_1^0$ at the top as the {\it bulk} spins, and 
$\sigma_j^n$ for arbitrary $j$ at the bottom as the {\it surface} spins. 

Pairwise Ising interaction is present between each neighboring spins connected by the line.
The Hamiltonian of the system is given by
\begin{equation}
H_n^{~}\left( \{ \sigma^{}_{} \} \right) = - J
\sum_{i = 0}^n
\sum_{j = 0}^{m(i) - 1} \sigma_j^i \sigma_{j+1}^i - J
\sum_{i = 0}^{n-1}
\sum_{j = 0}^{m(i)_{~}} \sigma_j^i \sigma_{2j}^{i+1} \, ,
\end{equation}
where $J > 0$ is the ferromagnetic coupling constant. In the left hand side, 
all the spins contained in the system is shortly denoted by $\{ \sigma \}$. 
We assume that there is no external magnetic field, unless otherwise noted. 
The thermodynamic properties of the system can be obtained from 
the partition function
\begin{equation}
Z_n^{~}( T ) = \sum_{\{ \sigma \}}^{~} \exp\left[ 
- \frac{H_n^{~}\left( \{ \sigma^{}_{} \} \right)}{k_{\rm B}^{~}T} \right]  \, ,
\end{equation}
where $T$ is the temperature, and where $k_{\rm B}^{~}$ is the Boltzmann constant. 
We set the temperature unit so that $k_{\rm B}^{~} = 1$ is satisfied. The sum of the 
Boltzmann weight of the whole system is taken for all the possible spin configurations. 

In this article, we perform numerical study on the system by the time evolving block decimation 
(TEBD) method~\cite{TEBD1,TEBD2} up to the case $n = 10$, and complementary by the 
modified CTMRG method for larger systems. We show that the surface spin expectation 
value at the center of the $n$-th row $\langle \sigma_{m(n)/2}^n \rangle$ is non-zero below 
$T_1^{~} = 1.58$, when $n$ is sufficiently large. On the other hand, the bulk spin expectation 
value $\langle \sigma_0^0 \rangle = \langle \sigma_1^0 \rangle$ becomes non-zero from
higher temperature $T_2^{~} = 2.269$. When $T$ is larger than $T_1^{~}$, the 
correlation function along the surface row 
shows power-law decay. 

The structure of this article is as follows. 
In Sec.~II, we shortly explain the way how to apply TEBD method, 
and show the calculated entanglement entropy and the correlation function. 
In Sec.~III we explain the numerical algorithm of the modified CTMRG method, 
which is complementary used for thermodynamic analysis, and show the calculated 
numerical results. Conclusions are summarized in the last section, 
and the remaining problems are discussed.

\section{Application of the TEBD Method}

In this section we explain how to perform the thermodynamic study on the Ising model
on the hierarchical pentagon lattice, by means of the TEBD method. Let us consider the 
distribution function 
\begin{equation}
G_n^{~}\left( \sigma_0^n, \cdots, \sigma_{m(n)}^n \right) = 
\sum_{\{ \sigma_{~}^0 \}}^{~} \cdots 
\sum_{\{ \sigma_{~}^{n-1} \}}^{~} \exp\left[ 
- \frac{H_n^{~}\left( \{ \sigma^{}_{} \} \right)}{k_{\rm B}^{~}T} \right] \, ,
\end{equation}
which represents the relative probability of appearance of the spin configuration $\sigma_0^n, 
\cdots, \sigma_{m(n)}^n$ at the bottom boundary. The configuration sum is taken for those 
row spins $\sigma_0^{i}, \cdots, \sigma_{m(i)}^{i}$, which are shortly denoted by 
$\{ \sigma^i_{~} \}$, from $i = 0$ to $i = n-1$. The left hand side can be written in the short 
form $G_n^{~}\left( \{ \sigma_{~}^{n} \} \right)$. Introducing the transfer matrix
\begin{eqnarray}
&& \!\!\!\!\!\!  U_i^{~}\left( \{ \sigma^{i+1}_{~} \} \, | \, \{ \sigma^i_{~} \} \right) = 
U_i^{~}\left( \sigma_0^{i+1}, \cdots \sigma_{m(i+1)}^{i+1} \, | \, 
\sigma_0^i, \cdots, \sigma_{m(i)}^i \right) \nonumber\\
&& \!\! = \exp\left[
\frac{J}{k_{\rm B}^{~} T} \!
\sum_{j = 0}^{m(i+1) - 1} \!\! \sigma_j^{i+1} \sigma_{j+1}^{i+1} + 
\frac{J}{k_{\rm B}^{~} T} 
\sum_{j = 0}^{m(i)_{~}} \sigma_j^i \sigma_{2j}^{i+1} 
\right] \, ,
\end{eqnarray}
we can obtain the distribution function in Eq.~(3) by way of the successive multiplication
of the transfer matrix 
\begin{equation}
G_{i+1}^{~}\left( \{ \sigma_{~}^{i+1} \}  \right) = 
\sum_{\{ \sigma_{~}^i \}}^{~} 
U_i^{~}\left( \{ \sigma^{i+1}_{~} \} \, | \, \{ \sigma^i_{~} \} \right) \, 
G_i^{~}\left( \{ \sigma_{~}^i \} \right) \, ,
\end{equation}
starting from the initial distribution 
\begin{equation}
G_0^{~}\left( \sigma_0^0, \sigma_1^0 \right) = 
\exp\left[ \frac{J}{k_{\rm B}^{~} T} \, \sigma_0^0 \, \sigma_1^0 \right] 
\end{equation}
at the top of the system. 
Figure 2 shows the pictorial representation of the transfer matrix multiplication in Eq.~(5) 
from $n = 0$ to $n = 2$. Configuration sums are taken for the spins shown by 
the black dots. Since $G_n^{~}\left( \{ \sigma_{~}^{n} \} \right)$ is the function of 
$m(n) + 1 = 2^n_{~} + 1$ number of the surface spins $\{ \sigma_{~}^{n} \}$, direct 
numerical calculation can be performed only up to several layers, around $n = 5$ or 
$n = 6$.

\begin{figure}
\begin{center}
\includegraphics[width = 5.0 cm]{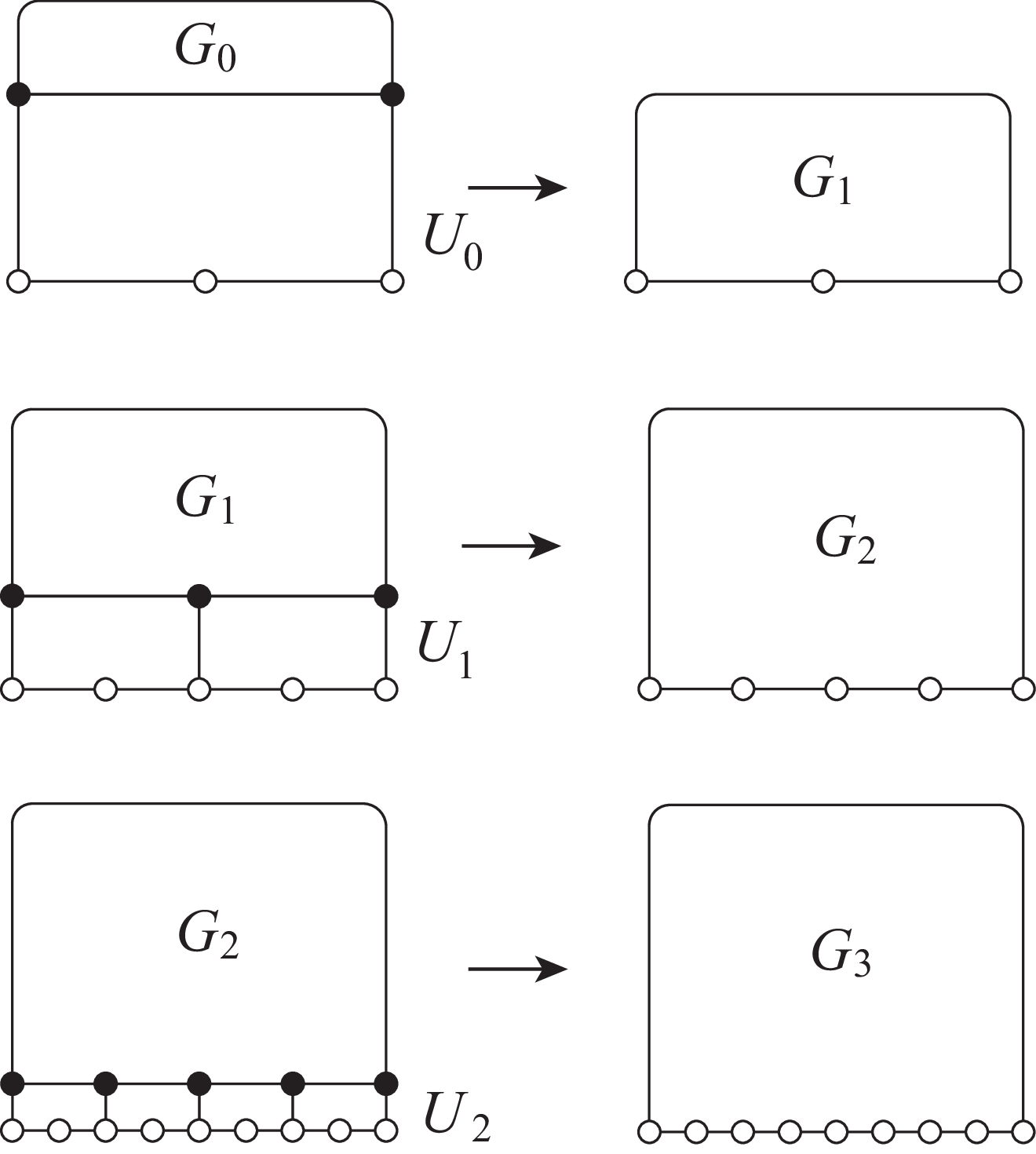}
\end{center}
\caption{Pictorial representation of the transfer matrix multiplications in Eq.~(5) 
from $i = 0$ to $i = 2$.}
\end{figure}

If we regard $G_n^{~}\left( \{ \sigma_{~}^{n} \} \right)$ as the quantum amplitude, the 
corresponding quantum state is expected to be weakly entangled, since the lattice can 
be horizontally separated by cutting only $n + 1$ horizontal bonds along the 
vertical cut, similar to the multi-scale entanglement renormalization Ansatz 
(MERA) network~\cite{MERA1,MERA2}. Thus $G_n^{~}\left( \{ \sigma_{~}^{n} \} \right)$ 
could be precisely represented by means of the matrix product state (MPS)~\cite{MPS1,MPS2}.
Since the transfer matrix in Eq.~(4) consists of horizontal product of local factors, the 
transfer matrix multiplication in Eq.~(5) can be efficiently performed step by step by means 
of the TEBD method~\cite{TEBD1,TEBD2}. 

We explain some details in the numerical transfer matrix multiplication, when the distribution 
function is represented in the form of  MPS. Those readers who are not interested in 
specific computational procedures can skip to the next subsection. In order to simplify the 
mathematical notations, we represent the Ising spins and corresponding tensor legs simply 
by alphabets, when the abbreviation is necessary~\cite{bit}. Suppose that we have 
the distribution function $G_2^{~}$ in the form of the mixed canonical MPS 
\begin{equation}
G_2^{~}( abcde ) = \sum_{\xi\mu}^{~}
L_{b}^{a\xi} \, D_{~}^{\, \xi} \, R_{~ \, c}^{\, \xi\mu} \, R_{~ \, d}^{\, \mu e} \, ,
\end{equation}
where we have expressed the row spin 
$\sigma_0^2, \sigma_1^2, \sigma_2^2, \sigma_3^2, \sigma_4^2$ simply by $abcde$. 
The 3-leg tensors $L_{b}^{a\xi}$, $R_{~ \, c}^{\, \xi\mu}$ and $R_{~ \, d}^{\, \mu e}$ satisfies 
the orthogonalities~\cite{MPS1,MPS2} 
\begin{eqnarray}
&& \sum_{ab}^{~} L_{b}^{a\xi} \, L_{b}^{a\xi'} = \delta_{\xi\xi'}^{~} \, , \nonumber\\
&& \sum_{\mu c}^{~} R_{~ \, c}^{\, \xi\mu} \, R_{~ \, c}^{\, \xi'\mu} = \delta_{\xi\xi'}^{~} \, , \nonumber\\
&& \sum_{de}^{~} R_{~ \, d}^{\, \mu e} \, R_{~ \, d}^{\, \mu' e} = \delta_{\mu\mu'}^{~} \, ,
\end{eqnarray}
and $D_{~}^{\, \xi}$ represents the singular value. In order to naturally arrange the spin 
indices in the equations, we put auxiliary indices on the upside of each tensors. 
Since all the tensor elements are real valued, we do not have to care about the complex 
conjugate. Hereafter we distinguish the 3-leg tensors and singular values by their indices.

\begin{figure}
\begin{center}
\includegraphics[width = 8.4 cm]{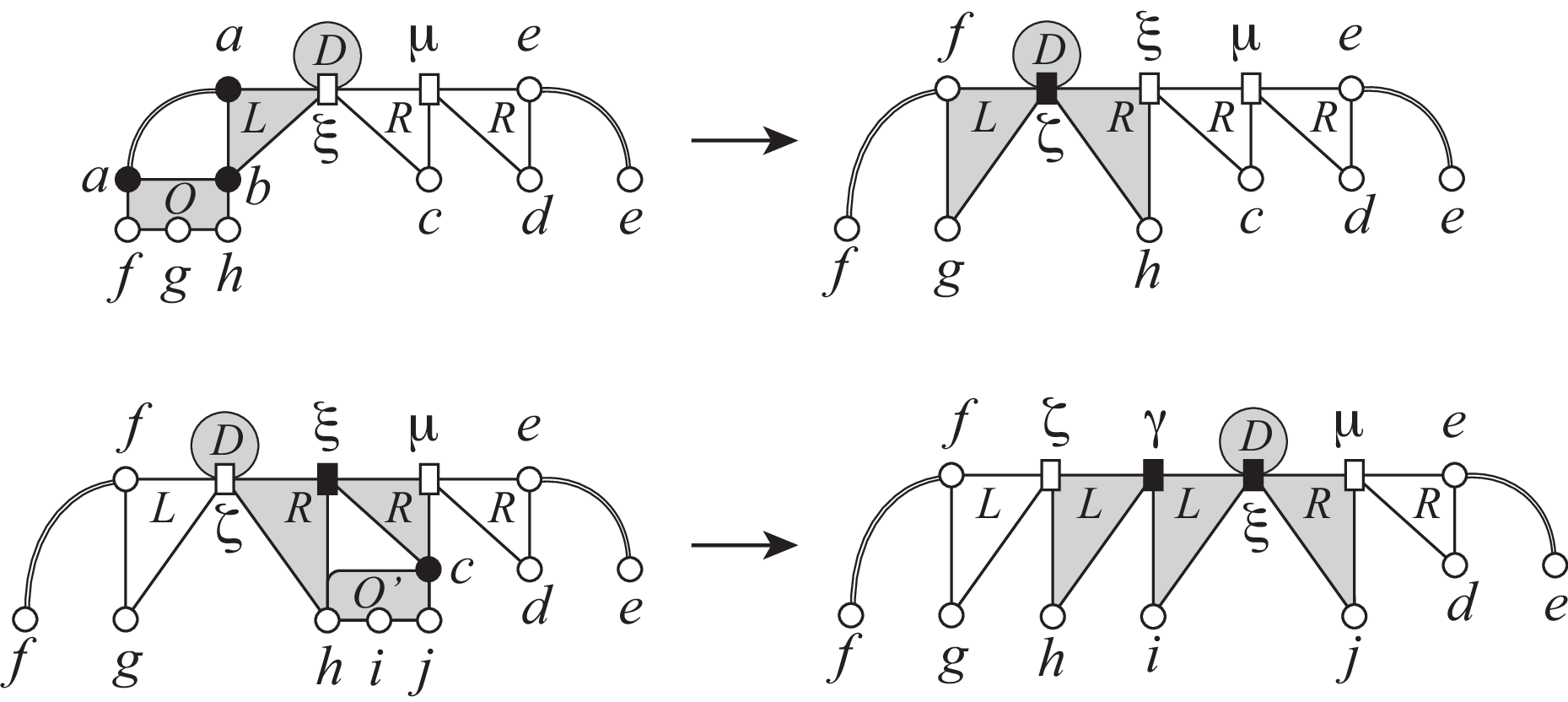}
\end{center}
\caption{Pictorial representation of the multiplication of $U_2^{~}$ to $G_2^{~}$ part by part, 
drawn by the IRF-type diagram. At the left and the right ends of the diagram, the spins 
connected by the arc are the same. }
\end{figure}

In the process of obtaining $G_3^{~}$ by multiplying $U_2^{~}$ to $G_2^{~}$, we 
perform the calculation part by part from left to right. We first prepare the 5-leg tensor
\begin{equation}
O^{\, a \,\,\, b}_{\, f g h} = \exp\left[ \frac{J}{k_{\rm B}^{~} T}( af + fg + gh + hb ) \right] \, ,
\end{equation}
which is a local factor contained in $U_2^{~}$, 
and perform the contraction with $L_{b}^{a\xi} \, D_{~}^{\, \xi}$ in the manner
\begin{equation}
X^{~~~\xi}_{fgh} = \sum_{ab}^{~} O^{\, a \,\,\, b}_{\, f g h} \, L_{b}^{a\xi} \, D_{~}^{\, \xi} \, .
\end{equation}
We then perform the singular value decomposition (SVD)  on $X^{~~~\xi}_{fgh}$, grouping 
the tensor legs to $fg$ and $h\xi$, to obtain the decomposed form 
\begin{equation}
X^{~~~\xi}_{fgh} = \sum_{\zeta}^{~} L^{f\zeta}_{g} \, D_{~}^{\, \zeta} \, R_{~ \, h}^{\, \zeta\xi} \, . 
\end{equation}
These local contraction and SVD processes are pictorially shown in the upper part 
of Fig.~3, where the whole part of the MPS is drawn in the form of the interaction round a 
face (IRF) type tensor-network diagrams~\cite{IRFdiagram}.

The next piece we multiply is the 4-leg tensor
\begin{equation}
{O'}^{~ \,\, c}_{\!\! h i j} = \exp\left[ \frac{J}{k_{\rm B}^{~} T}( hi + ij + jc ) \right] \, ,
\end{equation}
and we take the contraction in the manner
\begin{equation}
Y^{\zeta ~ \mu}_{hij} = \sum_{\xi c}^{~}
 {O'}^{~ \,\, c}_{\!\! h i j} \, D_{~}^{\, \zeta} \, R_{~ \, h}^{\, \zeta\xi} \, R_{~ \, c}^{\, \xi\mu} \, .
\end{equation}
This time we perform the SVD to $Y^{\zeta ~ \mu}_{hij}$ twice, and obtain the 
canonically decomposed form
\begin{equation}
Y^{\zeta ~ \mu}_{hij} = 
\sum_{\gamma\xi}^{~} L_{h}^{\zeta\gamma} \, L_{i}^{\gamma\xi} \, D_{~}^{\, \xi} \, R_{~ \, j}^{\, \xi\,\mu} 
\, .
\end{equation}
These processes are pictorially shown in the lower part of Fig.~3. In this manner, we can 
proceed to the next contraction with ${O'}^{~ \, d}_{\!\! j k \ell}$ and the following SVD, and 
also to the final contraction with ${O'}^{~ \,\,\, e}_{\!\! \ell m n}$ and the following SVD, 
to complete the transfer matrix multiplication $G_3^{~} = U_2^{~} \, G_2^{~}$. We finally 
obtain the MPS representation 
\begin{equation}
G_3^{~}( fghijk\ell mn ) = \sum_{\zeta\gamma\xi\mu\nu\rho}^{~}
L^{f\zeta}_{g} \, L_{h}^{\zeta\gamma} \, L_{i}^{\gamma\xi} \, L_{j}^{\xi\mu} \, L_{k}^{\mu\nu} \, 
L_{\ell}^{\nu\rho} \, D_{~}^{\, \rho} \, R_{~ m}^{\, \rho n} 
\end{equation}
of $G_3^{~}$,  where $D_{~}^{\, \rho}$ is located in the right side. Every time we perform 
SVD, we normalize the singular values so that their sum is unity. Thus in Eq.~(15), 
\begin{equation}
\sum_{\rho}^{~} D^{\, \rho}_{~} = 1 
\end{equation}
is satisfied in after the normalization. In the case we need $Z_n^{~}( T )$,
we store the normalization constant somewhere. 
It is convenient to move the singular value back to the left side, by means of successive 
reorthogonalization, in order to start the next multiplication $G_4^{~} = U_3^{~} \, G_3^{~}$ 
in the same manner as Eqs.~(9)-(16). 

It seems that the degree of freedom for $\zeta, \gamma, \xi, \mu, \nu,$ and $\rho$ in 
Eq.~(15) should be $4, 8, 16, 16, 8,$ and $4$ for the exact MPS representation of $G_3^{~}$, 
respectively, but actually the necessary singular values are $4, 8, 8, 8, 8,$ and $4$. 
The rank of $G_3^{~}$, when it is regarded as a matrix with respect to a specified 
bipartition of the spin row $fghijk\ell mn$, can be counted by considering how many 
interacting bonds should be cut at least in order to separate the whole lattice with $n = 3$ 
into two parts. In general, the rank of $G_n^{~}$ with respect to any bipartition is at 
most $2^n_{~}$, whereas there are $2^n_{~} + 1$ spins at the bottom. It is further 
possible to restrict the number of singular values in the numerical calculation by discarding 
tiny singular values, as it is commonly performed in tensor network applications.

\subsection{Calculated Results by the TEBD Method}

\begin{figure}
\begin{center}
\includegraphics[width = 6.6 cm]{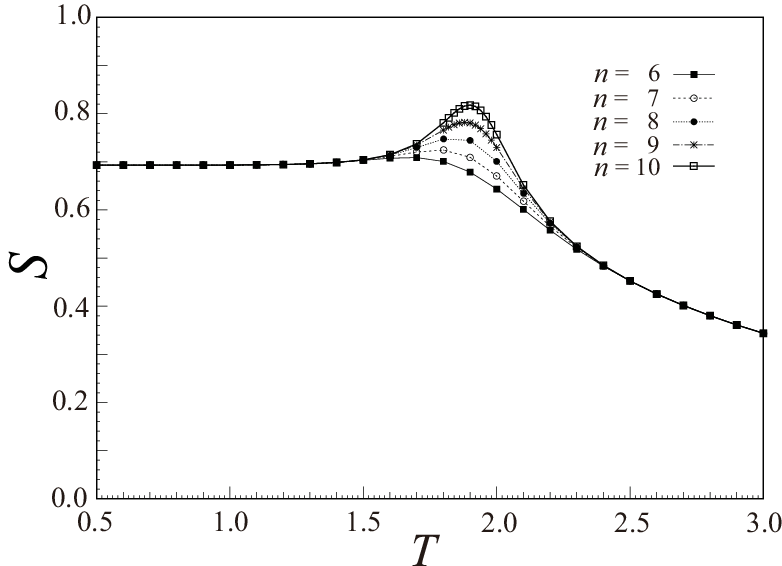}
\end{center}
\caption{Temperature dependence of the Entanglement entropy $S(T)$ in Eq.~(17). 
Curves are guide to eye.}
\end{figure}

We calculated the distribution function 
$G_n^{~}\left( \sigma_0^n , \cdots, \sigma_{m(n)}^n \right)$ up to $n = 10$ by means of
the TEBD method. The number of the kept singular values $\chi$ is automatically determined, 
so that the sum of the discarded singular values do not exceed the cut off parameter $\varepsilon$, 
where we vary it from $\varepsilon = 10^{-5}_{~}$ to $\varepsilon = 10^{-8}$ to confirm the 
numerical convergence of the obtained results. Hereafter we consider the case where the 
Ising interaction parameter $J$ in Eq.~(1) is unity.

We first focus on the division of the surface spin row $\{ \sigma^n_{~} \}$ to the left 
half $\sigma_0^n , \cdots \sigma_{m(n)/2}^{n}$ and the right half 
$\sigma_{m(n)/2+1}^{n} , \cdots, \sigma_{m(n)}^n$, and calculate the 
corresponding entanglement entropy (EE)
\begin{equation}
S( T ) = - \sum_{\kappa}^{~} D_{~}^{\, \kappa} \ln D_{~}^{\, \kappa} \, ,
\end{equation}
where $D_{~}^{\, \kappa}$ denotes the normalized singular value located between 
$\sigma_{m(n)/2}^{n}$ and $\sigma_{m(n)/2+1}^{n}$, when $G_n^{~}$ is represented 
in the form of MPS. Figure 4 shows the temperature dependence of $S( T )$ in the 
cases $n = 6, 7, 8, 9,$ and $10$. In low temperature we have $S( T ) = \ln 2$, which 
corresponds to the {\it superposition} of the complete ferromagnetic state, and in high 
temperature $S( T )$ monotonously decreases with $T$. There is a peak structure in 
the region $1.5 < T < 2.4$, where $n$-dependence of $S( T )$ is visible. The peak hight 
almost linearly increases with $n$, and therefore we conjecture that the hight diverges 
in the large $n$ limit at some temperature, which may be the surface transition 
temperature $T_2^{~} = 1.58$ that we will estimate in the next section. 

Next, let us observe the correlation function on the surface $\{ \sigma^{n=10}_{~} \}$. 
Figure 5 shows the correlation function
$\langle \sigma_{m(10) / 2 - \ell / 2}^{10} \, \sigma_{m(10) / 2 + \ell / 2}^{10} \rangle =
\langle \sigma_{512 - \ell / 2}^{10} \, \sigma_{512 + \ell / 2}^{10} \rangle$
with respect to the distance $\ell$, calculated in high temperature region where $S( T )$ is
converged with respect to $n$. As shown in the figure, power law decays with $\ell$ are 
observed, although there are minor fluctuation that arises from the inhomogeneous 
effect from the upper layers to the surface spin row. The presence of the power-law decay is 
in accordance with the Monte Carlo study by Asaduzzaman et. al. performed on finite 
hyperbolic $( 3, 7 )$ disks~\cite{Asaduzzaman}.

\begin{figure}
\begin{center}
\includegraphics[width = 6.5 cm]{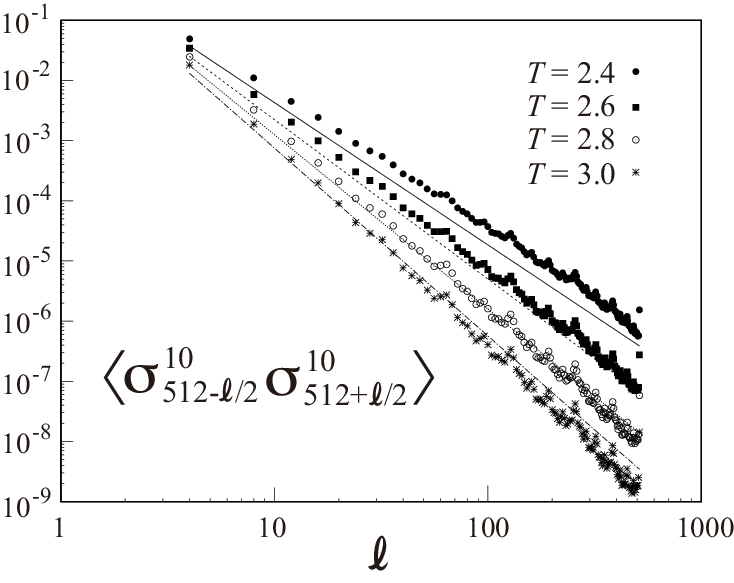}
\end{center}
\caption{Correlation function 
$\langle \sigma_{512 - \ell / 2}^{10} \, \sigma_{512 + \ell / 2}^{10} \rangle$ 
on the surface at $T = 2.4, 2.6, 2.8,$ and $3.0$. Lines denote the decay estimated from 
$\langle \sigma_0^n \sigma_{m(n)}^n \rangle$ calculated by the modified
CTMRG method through Eqs.~(33)-(35), for the corresponding temperatures.}
\end{figure}

\section{Application of the Modified CTMRG Method}

Complementary to the TEBD method, we introduce the modified CTMRG method, which can be 
used for the evaluation of the partition function $Z_n^{~}( T )$ in Eq.~(2) and spin expectation 
values, such as $\langle \sigma_0^0 \rangle$, 
$\langle \sigma_1^0 \rangle$, and $\langle \sigma_{m(n)/2}^n \rangle$, even when $n$ is 
relatively large. Those readers who are not interested in the numerical algorithm can 
skip to the next subsection.

The Boltzmann weight of the whole system, which appears in the r.h.s. of Eqs.~(2)-(3), 
can be represented as the product of the IRF weight 
$W^{\, a \,\, b}_{\, c d e}$ and the boundary weight $B_{fg}^{~}$, which are pictorially 
shown in Fig.~6. We use the IRF representation, since it is easy to treat Ising spins 
directly under the context of the (modified) CTMRG method~\cite{1978,Baxter}. 
The IRF weight is given by
\begin{equation}
W^{\, a \,\, b}_{\, c d e} = \exp\left[ \frac{J}{2 k_{\rm B}^{~} T}( ac + cd + de + eb + ba ) \right] \, ,
\end{equation}
where $a$, $b$, $c$, $d$, and $e$ represent the Ising spins on each vertex of the 
pentagon. Since each bond other than that on the system boundary is shared 
by adjacent pentagons, the parameter $J$ is divided by $2$ in the right hand side. 
Additionally we introduce the boundary weight
\begin{equation}
B_{fg}^{~} = \exp\left[ \frac{J}{2 k_{\rm B}^{~} T} \, fg \right] \, ,
\end{equation}
in order to adjust the Boltzmann weigh at the boundary of the system. Let us use the notation 
$B^{f}_{g}$ for the vertical boundary. In the case of the lattice with $n = 4$ in Fig.~1,
there are $2^n_{~} - 1 = 15$ IRF weights, $2n = 8$ vertical boundary weights, 
and $2^n_{~} + 1 = 17$ horizontal ones. 

\begin{figure}
\begin{center}
\includegraphics[width = 3.8 cm]{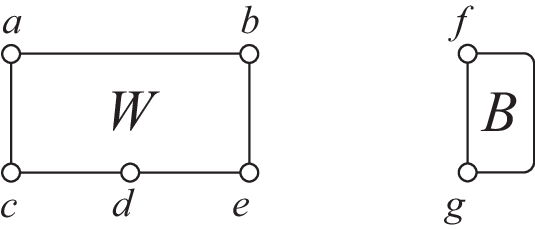}
\end{center}
\caption{Pictorial representation of the IRF weight $W^{\, a \,\, b}_{\, c d e}$ in Eq.~(18), 
and the boundary weight $B_{fg}^{~}$ in Eq.~(19). }
\end{figure}

In the CTMRG formulation, the whole system is divided into several components, and the 
Boltzmann weight for each component is calculated through the recursive area extensions 
and renormalization group transformations~\cite{1968,1978,CTMRG1,CTMRG2,Orus}. 
One of such weights on the hierarchical pentagon lattice are the series of the half-column 
transfer matrices (HCTM), which are located around the bottom of the system. The smallest one 
is given by
\begin{equation}
P_{\! 0 \,\, ce}^{\,\,\,\,ab} = \sum_d^{~} W^{\, a \,\, b}_{\, c d e} \, B_{cd}^{~} \, B_{de}^{~} \, ,
\end{equation}
where the position of the spins are shown in the upper side of Fig.~7. Boundary weights 
$B_{cd}^{~}$ and 
$B_{de}^{~}$ are multiplied, since $c$, $d$, and $e$ are spins on the bottom boundary, 
which is the surface. By definition, the HCTM satisfies the left-right symmetry 
$P_{\! 0 \,\, ce}^{\,\,\,\,ab} =
  P_{\! 0 \,\, ec}^{\,\,\,\,ba}$. Another series of the weights are the corner transfer matrices 
(CTM), which are located around the bottom left or the bottom right corners of the system. 
The smallest one around the bottom left is expressed as
\begin{equation}
C_{\! 0 \,\,\, ~e}^{\,\,\,\, ab} = 
\sum_{cd}^{~} B_{c}^{a} \, W^{\, a \,\, b}_{\, c d e} \, B_{cd}^{~} \, B_{de}^{~} = 
\sum_{c}^{~} B_{c}^{a} \, P_{\! 0 \,\, ce}^{\,\,\,\,ab} \, ,
\end{equation}
and similarly the one around the bottom right as
\begin{equation}
C_{\! 0 \,\, c~}^{\,\,\,\, ab} = 
\sum_{de}^{~} W^{\, a \,\, b}_{\, c d e} \, B_{cd}^{~} \, B_{de}^{~} B_{e}^{b} = 
\sum_{e}^{~} P_{\! 0 \,\,\, ce}^{\,\,\,\,ab} \, B_{e}^{b}  \, .
\end{equation}
For the latter convenience, let us use greek letters for those spins on the surface, in the 
manner as $P_{\! 0 \,\, \zeta\xi}^{\,\,\,\,ab}$, $C_{\! 0 \,\,\, ~\xi}^{\,\,\,\,ab}$ and 
$C_{\! 0 \,\, \zeta~}^{\,\,\,\,ab}$. 

\begin{figure}
\begin{center}
\includegraphics[width = 5.7 cm]{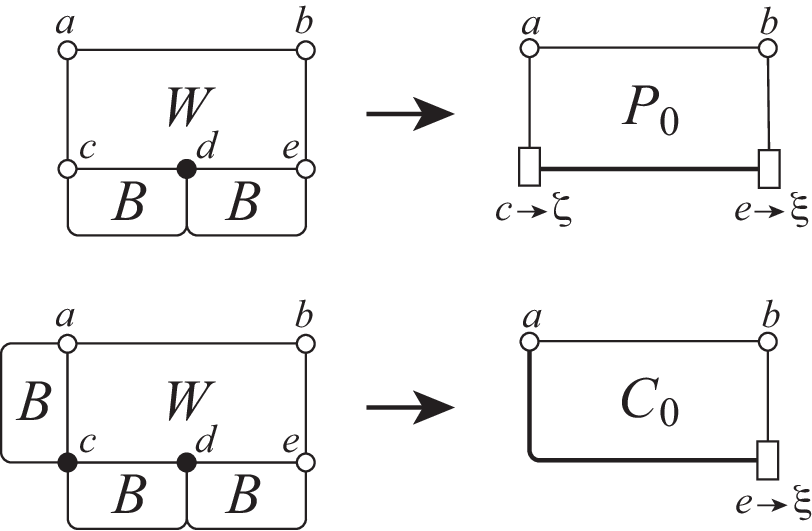}
\end{center}
\caption{The smallest HCTM $P_{\! 0 \,\, \zeta\xi}^{\,\,\,\,ab}$ in Eq.~(20) and the CTM
$C_{\! 0 \,\,\, ~\xi}^{\,\,\,\, ab}$ in Eq.~(21), which are located around the bottom of 
the system. Those contracted tensor legs are shown by the filled marks.}
\end{figure}

\begin{figure}
\begin{center}
\includegraphics[width = 5.7 cm]{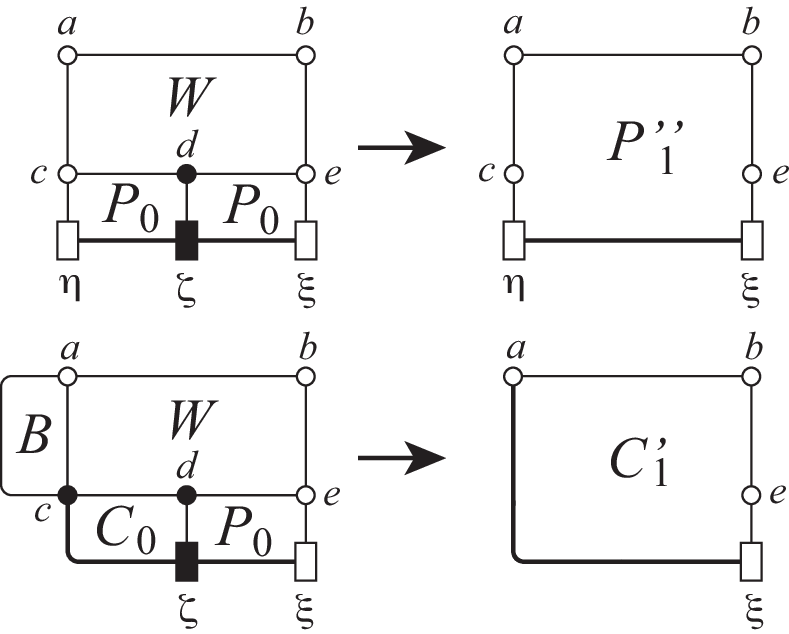}
\end{center}
\caption{The extension process of the HCTM in Eq.~(23), 
and that of the CTM in Eq.~(24).}
\end{figure}

The recursive structure of the hierarchical pentagon lattice enables the systematic extension
of the HCTM. Taking the contraction among $W$ and two $P_0^{~}$ in the manner
\begin{equation}
P''_{1}{ \begin{smallmatrix}  ab  \\ ce \\ \eta\xi  \end{smallmatrix}} = \sum_{d\zeta}^{~} 
W^{\, a \,\, b}_{\, c d e} \, P_{\! 0 \,\, \eta\zeta}^{\,\,\,\,cd} \, P_{\! 0 \,\, \zeta\xi}^{\,\,\,\,de} \, ,
\end{equation}
we obtain the extended HCTM. It should be noted that one can choose arbitrary letter for 
the tensor legs, since they are just the dummy indices that are used for the contractions 
among tensors. By definition, $P''_1$ satisfies the left-right symmetry 
$P''_{1}{ \begin{smallmatrix}  ab  \\ ce \\ \eta\xi  \end{smallmatrix}} = 
P''_{1}{ \begin{smallmatrix}  ba  \\ ec \\ \xi\eta  \end{smallmatrix}}$. 
Similar to Eq.~(23), the extension of the CTM at the bottom left corner is performed 
combining $W$, $C_0^{~}$, and $P_0^{~}$ as
\begin{equation}
C'_{1}{ \begin{smallmatrix}  ab  \\ ~e \\ ~\xi  \end{smallmatrix}} = \sum_{cd\zeta}^{~} B_c^a \, 
W^{\, a \,\, b}_{\, c d e} \, C_{\! 0 \,\, ~\zeta}^{\,\,\,\,cd} \, P_{\! 0 \,\, \zeta\xi}^{\,\,\,\,de} \, .
\end{equation}
Figure 8 pictorially represents the extension processes in Eqs.~(23) and (24). 
The extended CTM around the bottom right corner
$C'_{1}{ \! \begin{smallmatrix}  \,\, ab  \\ c~ \\ \eta ~  \end{smallmatrix}}$ can be obtained in 
the same manner, but we do not have to explicitly calculate it, since the left-right symmetry 
of the lattice allows us to use $C'_1$ in Eq.~(24) also for the bottom right corner, after the
appropriate substitution of indices. 

We have put dash marks on $P''_{1}$ and $C'_{1}$ in order to indicate that they have 
more tensor legs, respectively, compared with $P_{0}^{~}$ and $C_{0}^{~}$. It is better 
to represent the pair of legs $c$ and $\eta$, and also the pair $e$ and $\xi$ by something
like block spin variables. For this purpose, we first divide the legs of $P''_{1}$ to the pair 
$c\eta$ and the rest, and then perform SVD
\begin{equation}
P''_{1}{ \begin{smallmatrix}  ab  \\ ce \\ \eta\xi  \end{smallmatrix}} = \sum_{\mu}^{~} 
U^c_{\eta\mu} \, D_{\mu}^{~} \, 
V{ \begin{smallmatrix}  ab  \\ ~e \\ \mu\xi  \end{smallmatrix}} \, ,
\end{equation}
where $D_{\mu}^{~}$ denotes the singular values. 
We assume the decreasing order for $D_{\mu}^{~}$ with respect to $\mu$. 
The SVD we have performed is pictorially shown in the upper part of Fig.~9.

\begin{figure}
\begin{center}
\includegraphics[width = 5.7 cm]{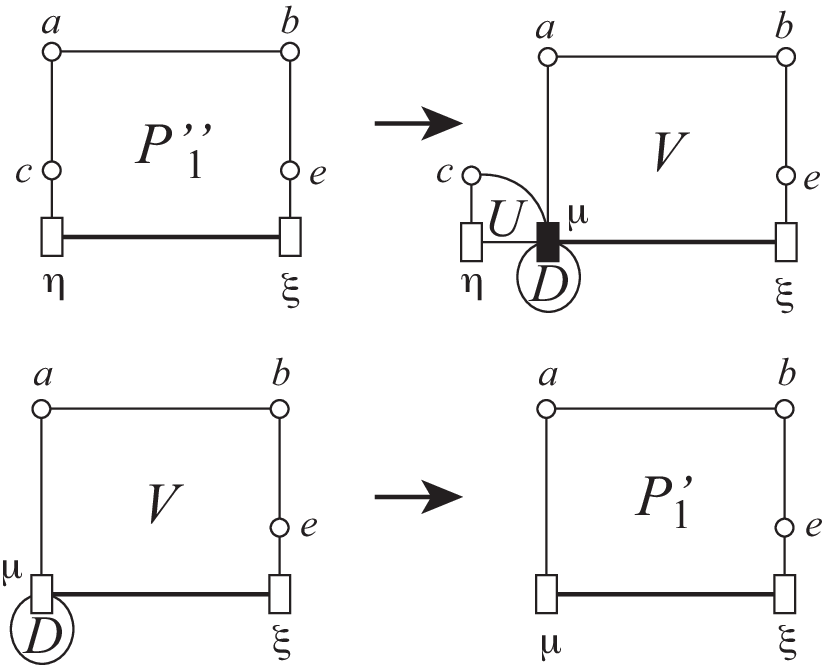}
\end{center}
\caption{The SVD in Eq.~(25) and the basis transformation applied to the left side of 
$P''_1$ in Eq.~(27).}
\end{figure}

The 3-leg tensor $U^c_{\eta\mu}$ in Eq.~(25) satisfies the orthogonality
\begin{equation}
\sum_{c\eta}^{~} U^c_{\eta\mu} \, U^c_{\eta\mu'} = \delta_{\mu\mu'}^{~} \, , 
\end{equation}
which enables us to use it as the basis transformation. 
Let us apply it on $P''_1$ in the manner
\begin{equation}
\sum_{c\eta}^{~} U^c_{\eta\mu} \,
P''_{1}{ \begin{smallmatrix}  ab  \\ ce \\ \eta\xi  \end{smallmatrix}} = 
\sum_{c\eta}^{~} U^c_{\eta\mu} \,
\sum_{\nu}^{~} 
U^c_{\eta\nu} \, D_{\nu}^{~} \, 
V{ \begin{smallmatrix}  ab  \\ ~e \\ \nu\xi  \end{smallmatrix}} =
D_{\mu}^{~} \, V{ \begin{smallmatrix}  ab  \\ ~e \\ \mu\xi  \end{smallmatrix}} \, 
\end{equation}
from the left side, 
and express $D_{\mu}^{~} V{ \begin{smallmatrix}  ab  \\ ~e \\ \mu\xi  \end{smallmatrix}}$ by
the new notation $P'_{1}{ \begin{smallmatrix}  ab  \\ ~e \\ \mu\xi  \end{smallmatrix}}$. 
We pictorially show the result of basis transformation in the lower part of Fig.~9. 
The left-right symmetry in $P''_{1}$ allows us to apply $U$ to the right side of
$P''_{1}$ and also to $P'_{1}$. For the latter, the transformation is performed as
\begin{equation}
\sum_{e\xi}^{~} P'_{1}{ \begin{smallmatrix}  ab  \\ ~e \\ \mu\xi  \end{smallmatrix}} \, U^e_{\xi\rho} 
= P_{\! 1 \,\, \mu\rho}^{\,\,\,\,ab} 
\end{equation}
to obtain the 4-leg tensor $P_{\! 1 \,\, \mu\rho}^{\,\,\,\,ab}$. The transformation $U$ can be 
applied to $C'_1$ in Eq.~(24) from the right side, since the lattice structure around the 
legs $b$, $e$, and $\xi$ is in common, as shown in Fig.~8. Performing the transformation
\begin{equation}
\sum_{e\xi}^{~} 
C'_{1}{ \begin{smallmatrix}  ab  \\ ~e \\  ~\xi  \end{smallmatrix}} \, U^e_{\xi\rho} =
C_{\! 1 \,\,\, ~\rho}^{\,\,\,\, ab} \, ,
\end{equation}
we obtain the 3-leg tensor $C_{\! 1 \,\,\, ~\rho}^{\,\,\,\, ab}$. 
Figure 10 pictorially shows the transformations in Eqs.~(28) and (29). 

\begin{figure}
\begin{center}
\includegraphics[width = 5.7 cm]{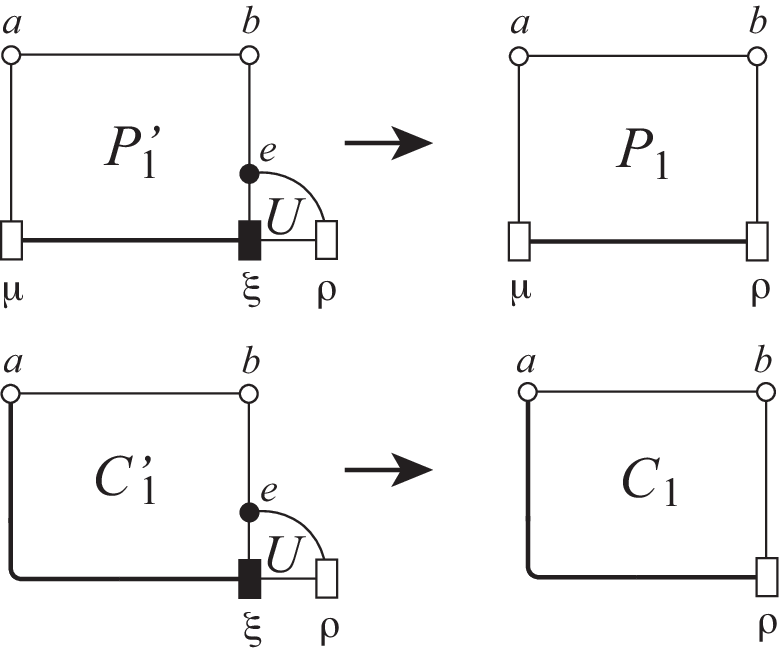}
\end{center}
\caption{The basis transformations applied to the right side of $P'_1$ in Eq.~(28), 
and that of $C'_1$ in Eq.~(29).}
\end{figure}

Initially we had $P_0^{~}$ in Eq.~(20) and $C_0^{~}$ in Eqs.~(21) and (22). 
Through the extension processes in Eqs.~(23) and (24) and the basis transformations 
in Eqs.~(27)-(29), we have obtained $P_1^{~}$ and $C_1^{~}$. 
It is possible to repeat the extensions and the transformations again to obtain 
$P_2^{~}$ and $C_2^{~}$. In this manner, we can construct the series of HCTMs 
$P_0^{~}$, $P_1^{~}$, $P_2^{~}$, $\cdots$, and that of CTMs 
$C_0^{~}$, $C_1^{~}$, $C_2^{~}$, $\cdots$. Every time we perform
the basis transformation, the degree of freedom of the greek indices becomes twice. 
This exponential increase of the freedom can be avoided by discarding tiny singular
values, and keeping only $\chi$ numbers of relevant basis in the transformations 
in Eqs.~(27)-(29), which can be regarded as the RG transformations. 

\begin{figure}
\begin{center}
\includegraphics[width = 3.7 cm]{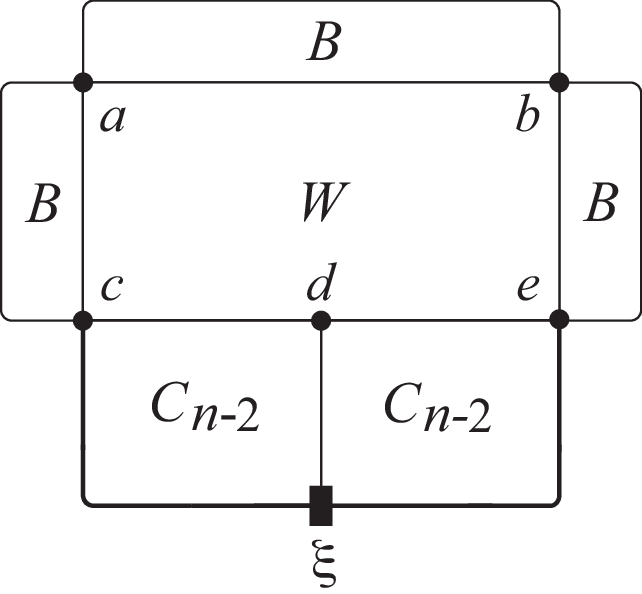}
\end{center}
\caption{The pictorial representation of $Z_{n}^{~}( T )$ in Eq.~(30).}
\end{figure}

We can obtain spin expectation values by means of the contraction of the tensors we
have obtained. First of all, the partition function in Eq.~(2) is calculated as
\begin{equation}
Z_n^{~}( T ) = \sum_{abcde\xi}^{~}
B_{ab}^{~} \, B^a_c \, W^{\, a \,\, b}_{\, c d e} \, B^b_e \, 
C_{\! n-2 \,\,\, ~\xi}^{~~~~~cd} \, 
C_{\! n-2 \,\,  \xi ~}^{~~~~~de} \, ,
\end{equation}
where the pictorial representation is shown in Fig.~11. The spin $a$ at the top is 
$\sigma_0^0$ in previous notation, thus its expectation value is expressed as
\begin{equation}
\langle \sigma_0^0 \rangle = \sum_{abcde\xi}^{~} \frac{a}{Z_n^{~}( T )} \, 
B_{ab}^{~} \, B^a_c \, W^{\, a \,\, b}_{\, c d e} \, B^b_e \, 
C_{\! n-2 \,\,\, ~\xi}^{~~~~~cd} \, 
C_{\! n-2 \,\,  \xi ~}^{~~~~~de} \, ,
\end{equation}
which is equal to $\langle \sigma_1^0 \rangle$. Rigorously speaking, the expectation
value $\langle \sigma_0^0 \rangle$ is zero when the number of layers $n$ is finite. Below the
symmetry breaking temperature $T_2^{~}$ of the bulk in the thermodynamic limit, however, 
numerically obtained $\langle \sigma_0^0 \rangle$ and  $\langle \sigma_1^0 \rangle$ 
becomes finite when $n$ is sufficiently large, since tiny numerical errors, which are 
common in floating point arithmetics, slightly break the spin inversion symmetry. Alternatively
we can impose a tiny external magnetic field $h$ to the spins on the surface $\{ \sigma_{~}^n \}$ 
to break the symmetry in a controlled manner. 

In case we do not need the value of $Z_n^{~}( T )$ directly, we can multiply arbitrary
factor to any tensors, since the factor cancels when we calculate the expectation 
values such as $\langle \sigma_0^0 \rangle$ in Eq.~(31). Often the normalization of
each tensor is performed so that the maximal absolute value of the element becomes 
unity, for the purpose of stabilize the floating-point arithmetics. 

\begin{figure}
\begin{center}
\includegraphics[width = 5.2 cm]{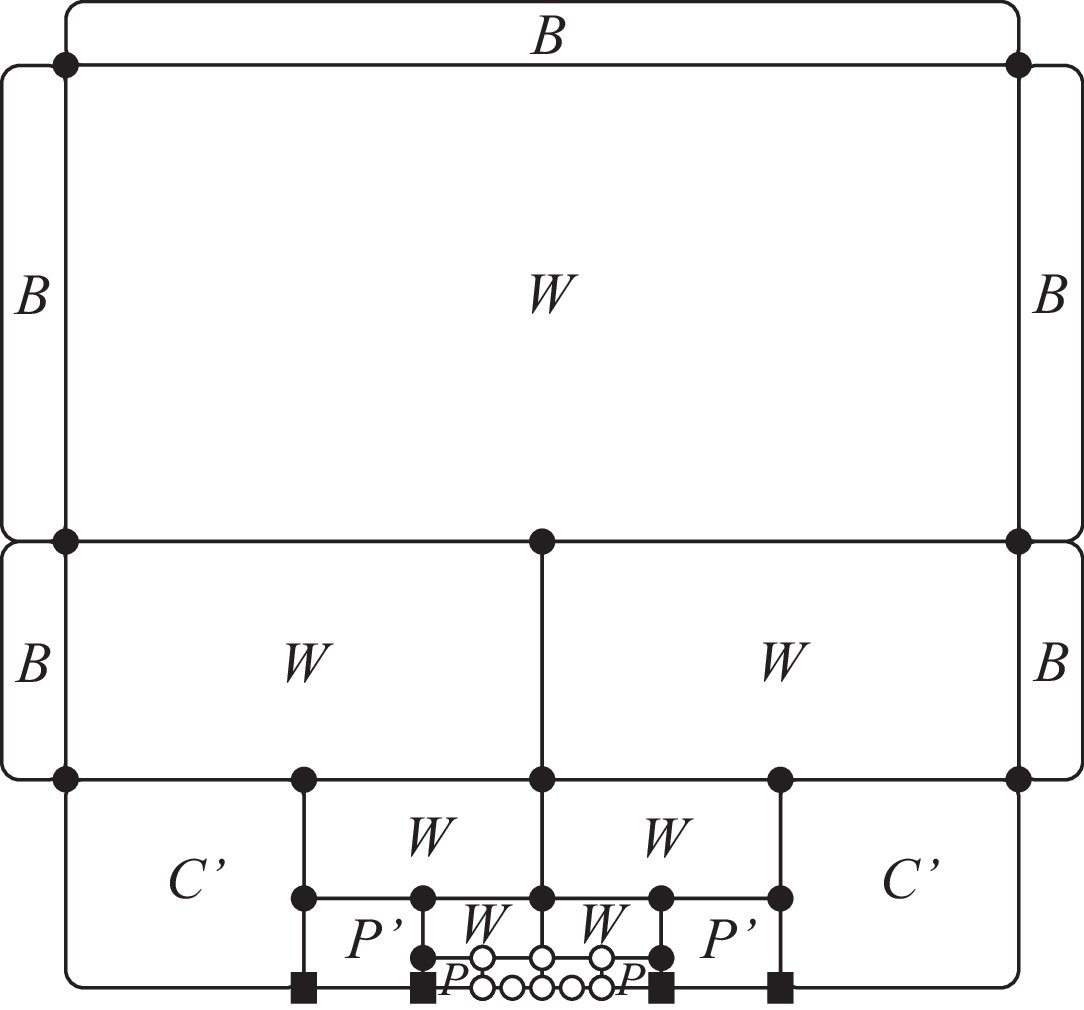}
\end{center}
\caption{The tensor $E^{a ~~ b ~~ c}_{defghi}$ used for the evaluation of 
$\langle \sigma^5_{n(5)/2} \rangle = \langle \sigma^5_{16} \rangle$. 
The tensor legs are denoted by open circles. }
\end{figure}

Combining all the tensors created up to the $n$-the iteration in the modified CTMRG 
algorithm, we can obtain the expectation value $\langle \sigma^n_{m(n)/2} \rangle$ at 
the center of the bottom row. Figure 12 shows the structure of the tensor 
$E^{a ~~ b ~~ c}_{defghi}$ that is necessary for the calculation of 
$\langle \sigma^5_{m(5)/2} \rangle = \langle \sigma^5_{16} \rangle$. 
The 8-leg tensor is created from the contraction
among $B$, $W$, $P_0^{~}$, $P'_1$, and $C'_2$, where all the tensor legs other 
than 8 legs shown by open circles on the lowest $W$ are summed up. It is important to 
take the configuration sum partially from up side to down side, as we performed in Eq.~(5), 
to suppress the computational time. The expectation value is then obtained as
\begin{equation}
\langle \sigma^5_{m(5)/2} \rangle = 
\frac{{\displaystyle \sum_{a \sim i}^{~}} \, f \, E^{a ~~ b ~~ c}_{defghi}}
{{\displaystyle \sum_{a \sim i}^{~}} \, E^{a ~~ b ~~ c}_{defghi}} =
\sum_{a \sim i}^{~} \frac{f}{Z_5^{~}( T )} \, E^{a ~~ b ~~ c}_{defghi} \, ,
\end{equation}
where the sum is taken over for all the legs. In the same manner, we can obtain 
$\langle \sigma^n_{m(n)/2} \rangle$ for arbitrary system size $n$. Also we can 
calculate the expectation values $\langle \sigma_{m(i)/2}^{i} \rangle$ from $i = 1$ to $n$, 
where $\sigma_{m( i )/2}^{i}$ are located vertically, arranging the tensors appropriately
and performing the contractions. Below the symmetry breaking temperature on the 
surface $T_1^{~}$, finite value is obtained for $\langle \sigma^n_{m(n)/2} \rangle$ 
when $n$ is sufficiently large.

It is also possible to obtain spin expectation values and correlation functions, 
creating another set of HRTMs and CTMs, each of which contain one of the IRF weights
$a \, W^{\, a \,\, b}_{\, c d e}$, $b \, W^{\, a \,\, b}_{\, c d e}$, $c \, W^{\, a \,\, b}_{\, c d e}$, 
$d \, W^{\, a \,\, b}_{\, c d e}$, and $e \, W^{\, a \,\, b}_{\, c d e}$
at the specified location. In this case, an Ising spin is implicitly contained in the renormalized
tensors. This approach is well known in the TRG formulations~\cite{TRG1,TRG2}, and 
we calculate the end-to-end correlation function $\langle \sigma_0^n \, \sigma_{m(n)}^n \rangle$ 
by this approach.

\subsection{Calculated Results by the Modified CTMRG Method}

\begin{figure}
\begin{center}
\includegraphics[width = 6.7 cm]{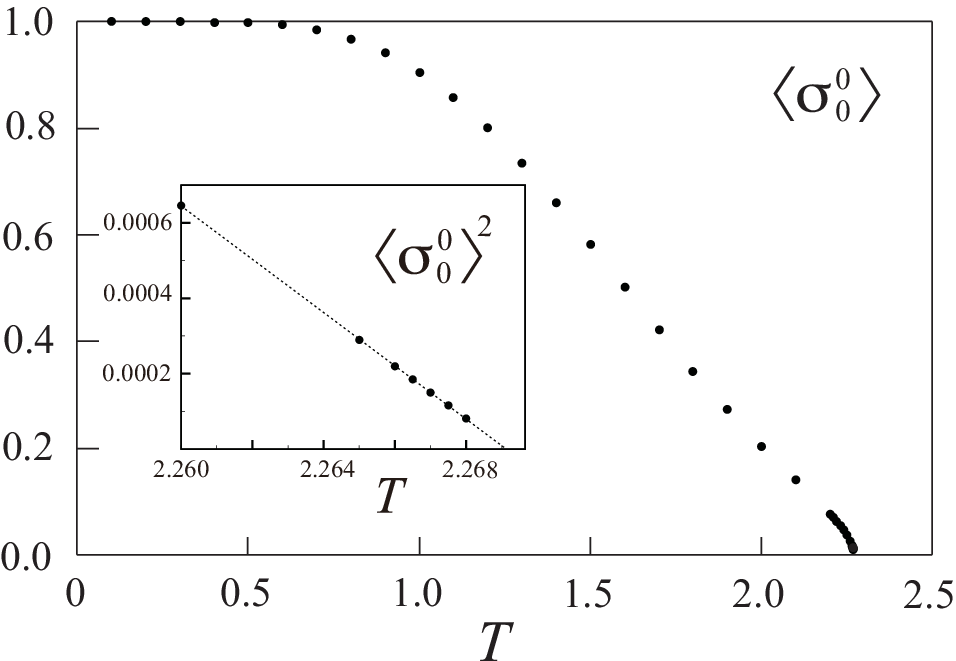}
\end{center}
\caption{Spontaneous magnetization $\langle \sigma_0^0 \rangle$ in the bulk.
The square $\langle \sigma_0^0 \rangle^2_{~}$ is shown in the inset.}
\end{figure}

We perform the numerical calculation on the hierarchical pentagon lattice by means of the 
modified CTMRG method, keeping $\chi = 55$ degrees of freedom at most when we 
perform the RG transformation. The expectation value $\langle \sigma_{m(n)/2}^n \rangle$ 
is converged with respect to $n$ less than $n = 100$ at any temperature. In case of 
$\langle \sigma_0^0 \rangle$, the convergence becomes slow near the transition 
temperature, and therefore we performed the iterative calculation up to $n = 30000$ 
at most. For all the data we show in this section, we checked the convergence with 
respect to $\chi$ and $n$. 

Figure 13 shows the temperature dependence of the spontaneous magnetization 
$\langle \sigma_0^0 \rangle$ at the top, which are regarded as the bulk part of the system. 
Around $T = 1$, the plotted value decreases with $T$, as if it vanishes some 
where between $T = 1.5$ and $T = 2.0$. But the decreasing rate becomes almost 
constant around $T = 1.5$, and finally the value vanishes at the transition temperature 
$T_2^{~} = 2.269$. As shown in the inset, $\langle \sigma_0^0 \rangle^2_{~}$ shows 
linear behavior near $T = T_2^{~}$. The behavior suggests that the transition is 
mean-field like, as it was observed in the bulk part of the hyperbolic 
lattices~\cite{Ueda,Krcmar,weak,Iharagi,Andrej_tri}. 

\begin{figure}
\begin{center}
\includegraphics[width = 6.7 cm]{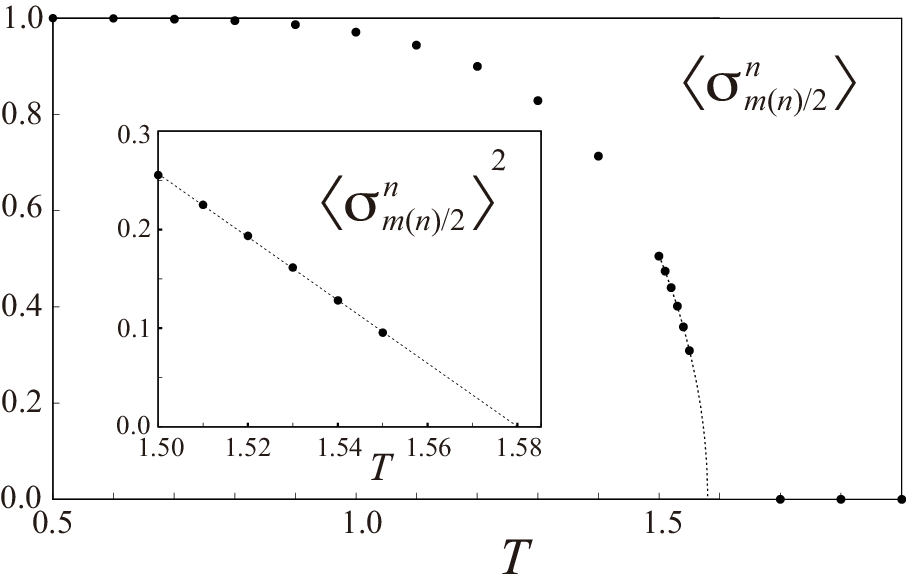}
\end{center}
\caption{Spontaneous magnetization $\langle \sigma_{m(n)/2}^{n} \rangle$ on the
surface, when $n$ is sufficiently large. The square 
$\langle \sigma_{m(n)/2}^{n} \rangle^2_{~}$ is shown in the inset.}
\end{figure}

Figure 14 shows the temperature dependence of $\langle \sigma_{m(n)/2}^{n} \rangle$ 
at the center of the bottom spin row, which are regarded as the surface of the system.
As shown in the inset, the squared value 
$\langle \sigma_{m(n)/2}^{n} \rangle^2_{~}$ shows linear behavior around the 
transition temperature $T_1^{~} = 1.58$. The dotted lines show the linear fitting result,
and the corresponding dotted curve is also drawn for $\langle \sigma_{m(n)/2}^{n} \rangle$. 
The transition is again mean-field like. The curious behavior of $\langle \sigma_0^0 \rangle$
in Fig.~13 might be related to presence of the symmetry breaking on the surface at $T_1^{~}$.

\begin{figure}
\begin{center}
\includegraphics[width = 6.0 cm]{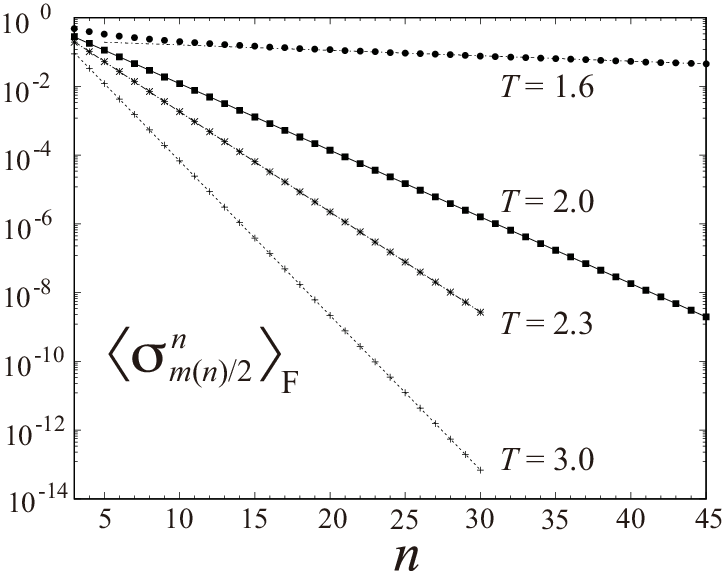}
\end{center}
\caption{Decay of $\langle \sigma_{m(n)/2}^{n} \rangle_{\rm F}^{~}$ with respect to $n$.}
\end{figure}

For the purpose of examining the value of $T_1^{~}$ by alternative view point, we observe 
the decay of spin correlation along the vertical direction. Let us impose the ferromagnetic 
boundary condition $\sigma_0^0 = \sigma_1^0 = 1$ at the top of the system, when we 
calculate the environment tensor $E^{a ~~ b ~~ c}_{defghi}$ shown in Fig.~12. We denote 
the corresponding expectation value as $\langle \sigma^n_{m(n)/2} \rangle_{\rm F}^{~}$. 
Below $T_1^{~}$, we obtain the same value 
$\langle \sigma^n_{m(n)/2} \rangle = \langle \sigma^n_{m(n)/2} \rangle_{\rm F}^{~}$ as 
the spontaneous magnetization, when $n$ is sufficiently large. 
Above $T_1^{~}$, $\langle \sigma^n_{m(n)/2} \rangle_{\rm F}^{~}$ 
show exponential dumping with respect to $n$, as shown in Fig.~15. The dotted lines denote
the linear fitting result in the region where the exponential dumping 
\begin{equation}
\langle \sigma_{m(n)/2}^{n} \rangle_{\rm F}^{~} \propto e^{- b(T) n}_{~}
\end{equation}
is observed clearly, 
where $b( T )$ is the decay rate. Only in the case $T = 1.6$, several plots visibly deviate from 
the dotted line in the small $n$ region. Figure 16 shows the temperature dependence of $b( T )$. 
No singular behavior is observed at the bulk transition temperature $T_2^{~} = 2.269$, and 
$b( T )$ almost linearly decreases to zero at $T_1^{~} = 1.58$. The result may suggest that the spin 
correlation length in the vertical direction diverges at $T_1^{~}$ with the critical index $\nu = 1$, 
which is different from the mean-field value $\nu = 1/2$. 
It should be noted, however, that the lattice is not uniform in the vertical direction, 
and therefore the definition of the effective distance in vertical direction is not 
straightforward, when we consider continuum limit at $T_1^{~}$. 

\begin{figure}
\begin{center}
\includegraphics[width = 6.5 cm]{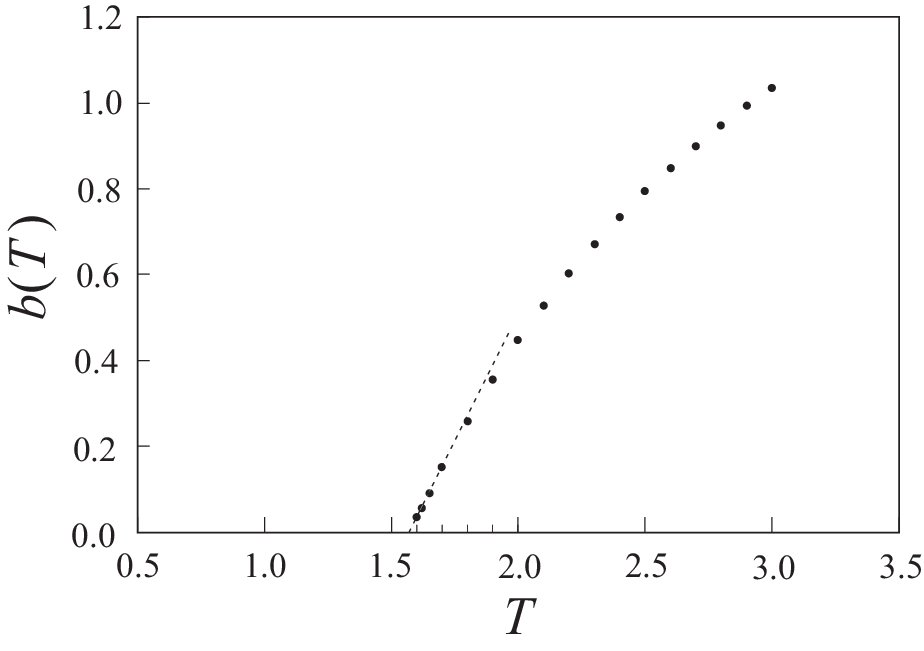}
\end{center}
\caption{Decay rate $b( T )$ in Eq.~(33) with respect to $T$. 
The dotted line connects the plots at $T = 1.6$ and $T = 1.7$.}
\end{figure}

In the intermediate temperature region $T_1^{~} < T < T_2^{~}$, the bulk spin has finite 
spontaneous magnetization as shown in Fig.~13, and the surface magnetization is absent
as shown in Fig.~14. In order to capture the magnetization profile between these two parts, 
we calculate the expectation values of the vertically aligned spins 
$\langle \sigma_{m(i)/2}^{i} \rangle$ from  $i = 1$ to $i = 150$ in the case where the system 
contains $n = 150$ layers. In order to observe the symmetry breaking in a controlled manner, 
we impose a weak magnetic field $h$ to the surface spins $\{ \sigma^n_{~} \}$. The initial
HCTM is then modified as
\begin{equation}
P_{\! 0 \,\, ce}^{\,\,\,\,ab} = \sum_d^{~} W^{\, a \,\, b}_{\, c d e} \, B_{cd}^{~} \, B_{de}^{~} \, 
\exp\left[ \frac{K}{2} ( c + 2d + e ) \right] \, ,
\end{equation}
where the parameter $K = \mu_{\rm B}^{~} h / k_{\rm B}^{~} T$ represents the effect of the
magnetic interaction with the Bohr moment $\mu_{\rm B}^{~}$. Similarly, the initial CTM is
modified as
\begin{equation}
C_{\! 0 \,\,\, ~e}^{\,\,\,\, ab} = 
\sum_{cd}^{~} B_{c}^{a} \, W^{\, a \,\, b}_{\, c d e} \, B_{cd}^{~} \, B_{de}^{~} 
\exp\left[ \frac{K}{2} ( 2 c + 2d + e ) \right]
\, .
\end{equation}
We perform the calculation for the cases where $K = 10^{-4}_{~}, 10^{-6}_{~}, 10^{-8}_{~}, 
10^{-10}_{~}$, and $0$. Figure~17 shows the calculated $\langle \sigma_{m(i)/2}^{i} \rangle$ 
from $i = 5$ to $i = 150$ when $T = 1.8$. Exponential dumping with respect to $i$ is clearly 
observed near the surface, and the dumping rate is consistent with $b( T = 1.8 )$ obtained 
from the plots at $T = 1.8$ in Fig.~15. In the case $K = 0$, the surface magnetization 
$\langle \sigma_{m(150)/2}^{150} \rangle$ is {\it artificially induced} by a tiny numerical error. 
The magnetic profiles plotted in Fig.~17 shows that there is a polarized area in the bulk part. 
The situation is common to the Ising model on the Cayley tree below the bulk
symmetry breaking temperature.

\begin{figure}
\begin{center}
\includegraphics[width = 6.2 cm]{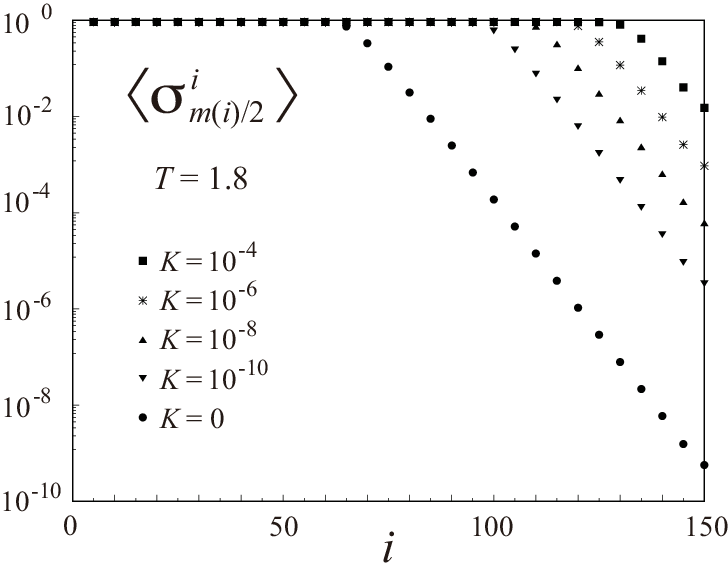}
\end{center}
\caption{The expectation values $\langle \sigma_{m(i)/2}^i \rangle$ from $i = 5$ to 
$i = 150$ that are calculated for the system with $n = 150$ layers when $T = 1.8$.}
\end{figure}

\begin{figure}
\begin{center}
\includegraphics[width = 6.2 cm]{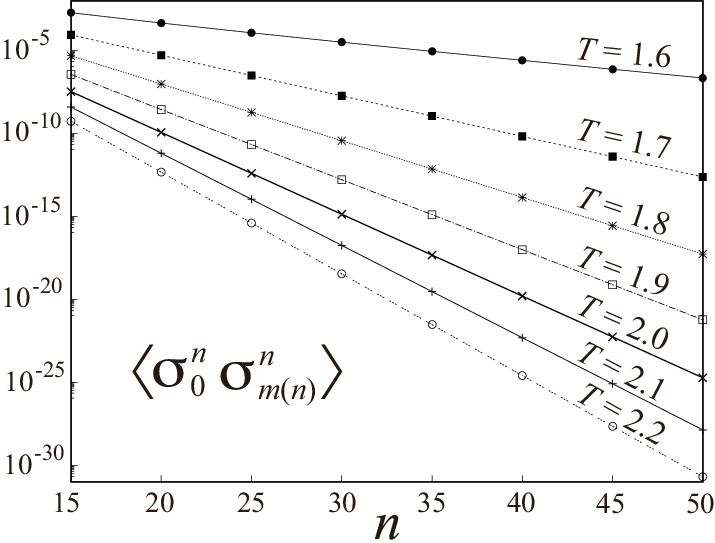}
\end{center}
\caption{Correlation function $\langle \sigma_0^n \sigma_{m(n)}^n \rangle$.}
\end{figure}

Finally, let us examine how strongly $\sigma_0^n$ and $\sigma_{m(n)}^n$ are correlated, 
observing the expectation value $\langle \sigma_0^n \sigma_{m(n)}^n \rangle$ for the zero field 
case $K = 0$. Figure 18 shows the calculated result from $T = 1.6$ to $T = 2.2$ by the step 
$\Delta T = 0.1$, with respect to $n$. The exponential dumping 
\begin{equation}
\langle \sigma_0^n \sigma_{m(n)}^n \rangle \propto e^{- c(T) n}_{~}
\end{equation}
is clearly observed, where $c( T )$ is the dumping constant. It should be noted that the 
distance $\ell$ between $\sigma_0^n$ and $\sigma_{m(n)}^n$ measured along the surface 
is $m(n) = 2^n_{~}$. From the relation
\begin{equation}
\ell^{-\eta}_{~} = \left( 2^n_{~} \right)^{-\eta}_{~} \propto e^{- c(T) n}_{~} \, ,
\end{equation}
the exponent for the power-law decay along the surface is estimated as 
\begin{equation}
\eta = \frac{c(T)}{\ln 2} \, .
\end{equation}
Figure 19 shows the temperature dependence of $c( T )$. 
Considering the fact that the shortest path from $\sigma_0^n$ to $\sigma_{m(n)}^n$ is
$2n + 1$, it is expected that $c( T )$ in Eq.~(36) is nearly twice as large as $b( T )$ in 
Eq.~(33). Comparing Fig.~16 and Fig.~19, one finds that the relation 
$c( 3.0 ) \sim 2 b( 3.0 )$ is satisfied, where the correlation length to the vertical 
direction is short at this temperature. 
From the value of $c( T )$ at $T = 2.4, 2.6, 2.8,$ and $3.0$, we calculate $\eta$ 
by Eq.~(38), and draw the corresponding lines in Fig.~5. Qualitative agreement is 
observed about the decay rate estimated from the TEBD method and that from the 
modified CTMRG method.
Near $T = T_1^{~}$, where the vertical correlation length is long, the relation 
$c( T ) \sim 2 b( T )$ does not hold any more. The values $\left[ c( T ) \right]^{1.6}_{~}$ 
neat $T = T_1^{~}$ shown in the inset are nearly proportional to $T - T_1^{~}$. 
If we draw the line that passes the plots $\left[ c( 1.6 ) \right]^{1.6}_{~}$ and 
$\left[ c( 1.7 ) \right]^{1.6}_{~}$ in the inset, we obtain $T_1^{~} = 1.56$, where
the last digit changes if we use $\left[ c( T ) \right]^{1.5}_{~}$ or $\left[ c( T ) \right]^{1.7}_{~}$, 
and thus we conjecture that $T_1^{~} = 1.58$ estimated from Fig.~16 is more accurate. 

\begin{figure}
\begin{center}
\includegraphics[width = 6.2 cm]{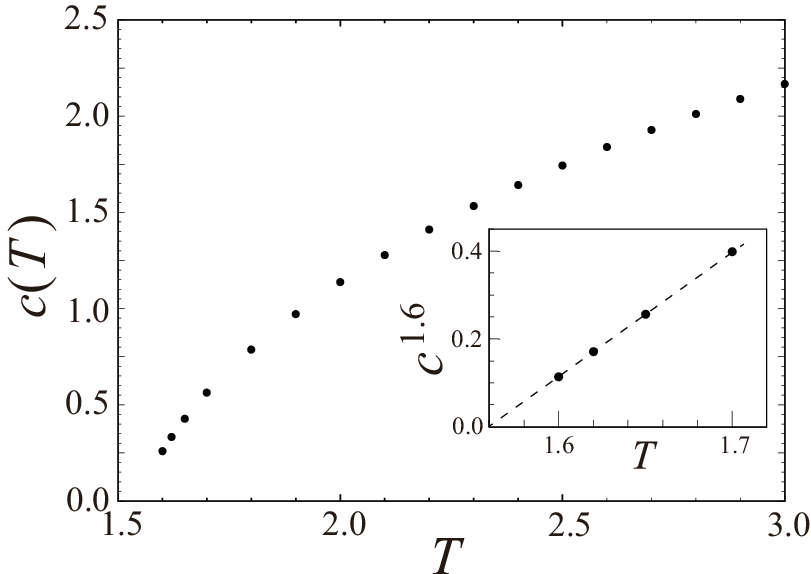}
\end{center}
\caption{Temperature dependence of the exponent $c( T )$ in Eq.~(36).
We show $\left[ c( T ) \right]^{1.6}_{~}$ neat $T = T_1^{~}$ in the inset.}
\end{figure}

\section{Conclusions and Discussions}

We numerically observed the phase transition of the Ising model on the hierarchical pentagon
lattice, by means of the tensor network methods. By means of the TEBD method up to the 
number of layers $n = 10$, the distribution function $G_n^{~}$ on the surface is calculated. 
The corresponding entanglement entropy $S( T )$ with respect to the bipartition of the 
surface spin row shows peak structure, whose hight increases with $n$. In the high 
temperature region $T \ge 2.4$, the power-law decay of the correlation function along the 
surface is observed. By means of the modified CTMRG method, the bulk transition of the 
mean-field type is detected at $T_2^{~} = 2.269$, at the top of the system. On the surface, 
which is the spin row at the bottom, the order-disorder transition of the mean-field type is 
detected at $T_1^{~} = 1.58$, which is lower than $T_2^{~}$. The end-to-end correlation 
function $\langle \sigma_0^n \sigma_{m(n)}^n \rangle$ shows exponential dumping with
$n$ above  $T_1^{~}$, which agrees with the power-law decay of the spin correlation 
function observed by the TEBD method. 

It should be noted that the behavior of the inverse of the vertical correlation length 
shown in Fig.~16 does not directly matches the mean-field behavior of the spontaneous 
magnetization on the surface near $T = T_1^{~}$. The fact reminds us the non uniformity 
of the lattice in the vertical direction, and it is not obvious that one can naively introduce
the scaling relation on this lattice structure, in particular when we consider the continuum limit 
in the neighborhood of $T = T_1^{~}$. To solve the puzzle is one of our future study.

Contrary to the thermodynamics on the Cayley tree, we observed the phase transition on 
the surface, which is the outer system boundary, below $T_1^{~}$. This difference can be 
explained by the presence of loops in hierarchical pentagon lattice. It should be noted that the 
surface spin row $\{ \sigma^n_{~} \}$ can be regarded as the one-dimensional Ising model, 
where upper layers effectively induce long-range interactions. The similar structure is present
also in hyperbolic $( p, q )$ lattices, and therefore phase transition could be present on the 
boundary in these  hyperbolic lattices.

The hierarchical pentagon lattice we treated in this article can be considered 
as a fractal lattice, in the sense that it has self similarity. On the fractal lattice 
such as the Sierpinski carpet, it is known that the critical behavior is highly 
dependent on the location of the site in the system~\cite{fractal1,fractal2}. 
Thus the possible coming study is to observe spin expectation values 
$\langle \sigma_j^i \rangle$ and correlation functions 
$\langle \sigma_j^i \sigma_k^{\ell} \rangle$ for a various combination of $i, j, k,$ and $\ell$. 
In principle, it is at least possible to target arbitrary pair of spins from the row spins 
$\{ \sigma^n_{~} \}$ on the surface, and obtain expectation values such as 
$\langle \sigma_j^n \rangle$ and $\langle \sigma_j^n \sigma_{\ell}^n \rangle$
for arbitrary $j$ and $\ell$. To construct a systematic numerical algorithm, 
which automatically choose the necessary pieces of tensors for the targetted
spins, is one of the next computational challenge in the modified CTMRG method, 
which has many aspects in common with the tensor renormalization group 
(TRG)~\cite{TRG1,TRG2} studies. 


In the application of the TEBD method, every time we multiply the transfer matrix, the 
number of 3-leg tensor contained in the MPS representation of $G_n^{~}$ becomes 
almost twice. This is the reason why our TEBD calculation is limited up to $n = 10$. 
In the case when we are only interested in the region around the center of the surface spin row, 
we can ignore those 3-leg tensors that are located near the left and right boundary of 
the row, and can {\it shrink} the length of MPS. Such an approximation is possible 
in the case where the surface area extension is more rapid than the propagation of 
correlation effect. Such a numerical trick is similar to the tensor eliminations in the co-moving 
MPS window method~\cite{comoving1,comoving2}, performed in the back side of the window.

\begin{acknowledgment}

T.N. was supported by JSPS KAKENHI Grant Number 21K03403 and by the COE research 
grant in computational science from Hyogo Prefecture and Kobe City through Foundation for 
Computational Science. We thank to valuable discussions with K.~Okunishi and H.~Ueda.
\end{acknowledgment}


\begin{thebibliography}{99}
\bibitem{bit} When we consider the tensor legs $a, b, c, d$ and $e$ as the Ising spins, 
they take the value either $1$ or $-1$. In the computational programing, they are treated 
as bits, that take either $0$ or $1$.
\end{thebibliography}
\end{document}